\documentclass[10pt]{article}
\usepackage[compress]{cite}
\usepackage[utf8]{inputenc}
\usepackage{mathtools}
\usepackage{graphicx}
\usepackage{color}
\begin{document}

\markboth{P.Cunha,  C. Herdeiro, E. Radu \& H. Rúnarsson}
{Shadows of Kerr Black Holes with scalar hair }

\title{Shadows of Kerr black holes with\\ and without scalar hair }

\author{\footnotesize Pedro V. P. Cunha\footnote{\footnotesize\textit{Departamento de Física da Universidade de Aveiro and Center for Research and Development in Mathematics and Applications (CIDMA), Campus de Santiago, 3810-183 Aveiro, Portugal.}} \footnote{\footnotesize\textit{CENTRA,Departamento de Física, Instituto Superior Técnico, Universidade de Lisboa, Avenida Rovisco Pais 1, 1049 Lisboa, Portugal.}} \hspace{1mm}(\textit{cunhapcc@gmail.com}),\\ \footnotesize Carlos A. R. Herdeiro$^*$ (\textit{herdeiro@ua.pt}),\\\footnotesize Eugen Radu$^*$ (\textit{eugen.radu@ua.pt}),\\\footnotesize Helgi F. Rúnarsson$^*$ (\textit{helgi.runarsson@gmail.com}) 
}

\date{}
\maketitle

\begin{abstract}
For an observer, the Black Hole (BH) shadow is the BH's apparent {image} in the sky due to the gravitational lensing of nearby radiation, emitted by an external source. A recent class of solutions dubbed \textit{Kerr BHs with scalar hair} possess smaller shadows than the corresponding Kerr BHs and, under some conditions, novel exotic shadow shapes can arise. Thus, these hairy BHs could potentially provide new shadow templates for future experiments such as the Event Horizon Telescope. In order to obtain the shadows, the backward ray-tracing algorithm is briefly introduced, followed by numerical examples of shadows of Kerr BHs with scalar hair contrasting with the Kerr analogues. Additionally, an analytical solution for the Kerr shadow is derived in closed form for a ZAMO observer at an arbitrary position.   
\end{abstract}




\section{Introduction}
The Kerr space-time\cite{Kerr} {is, according to the uniqueness theorems \cite{Uniqueness}, the only stationary, regular, asymptotically flat BH solution of vacuum General Relativity. As such, it is believed to describe the endpoint of gravitational collapse when (essentially) all matter has either fallen to the BH or been scattered towards infinity.} However, in order to provide alternative shadow templates for upcoming astronomical observations, it is timely to consider different types of solutions {(namely non-vacuum)}. In particular, templates are necessary for future observations of the supermassive BH candidate Sgr A* at the center of our galaxy with the Event Horizon Telescope \cite{EHT,Johannsen_review}. 
Hence, a novel class of BHs in equilibrium with a massive scalar field, \textit{Kerr BHs with scalar hair}, will be considered. This class exhibits various intriguing properties, including shadows that differ significantly from the Kerr prediction \cite{Cunha, Cunha_master}. \\
{The \textit{shadow}\cite{Bardeen,Johannsen} is the region in the observer sky corresponding to the incidence directions of photons that originate asymptotically from the event horizon. It is the optical perception of a BH by an observer, contrasting over the background light.} The shadow can be associated with the BH's light absorption cross-section at high frequencies, and its {outline} in the sky depends on the gravitational lensing of nearby radiation{, thus bearing the fingerprint of the geometry around the BH \cite{Johannsen}.}\\ 
Although in some cases (e.g. Kerr) it is possible to have an analytical closed form for the shadow edge, in general this is not possible and numerical methods are required. For the latter case, the evolution of light rays is performed numerically by solving the null geodesics equation. The general form of the geodesic equations is given by \cite{Novikov}: 
\begin{equation}\ddot{x}^\mu+\Gamma^\mu_{\alpha\beta}\,\dot{x}^\alpha\dot{x}^\beta=0,\end{equation}
where the derivative is taken with respect to an affine parameter and $\Gamma^\mu_{\alpha\beta}$ are the Christoffel symbols. This system of four second order differential equations can be simplified if symmetries of the space-time exist, with some of the four equations reduced to first order.\\
Numerically, the most naive approach would be to evolve the light rays directly from the source and detect which ones reach the observer. However this procedure is very inefficient since most rays would not reach it and thus spend unnecessary computation time. A better approach is to evolve the light rays from the observer backward in time and identify their origin \cite{Interstellar_article1}, a method named \textit{backward ray-tracing}. The information carried by each ray would then be respectively assigned to a pixel in a final image, which embodies the optical perception of the observer.\\
This paper is divided in three main sections; in the first section the backward ray-tracing setup is discussed; the following section focuses on the Kerr space-time and the last section addresses the case of Kerr BHs with scalar hair. Natural units $G=c=1$ will be used.\\

\section{Backward Ray-tracing}
\subsection{Local observer basis}
In this section the space-time is assumed stationary and axially symmetric, and thus having two Killing vector fields. Additionally, it is also assumed to be asymptotically flat. The notation for the coordinate set $\{t,r,\theta,\varphi\}$ was chosen in order to appear spherical-like. In particular, $t$ is the time coordinate and $\varphi$ the azimuthal coordinate, each adapted to the corresponding Killing vector.\\  
The observer basis $\{\hat{e}_{(t)},\hat{e}_{(r)},\hat{e}_{(\theta)},\hat{e}_{(\varphi)}\}$ can be expanded in the coordinate basis $\{\partial_t,\partial_r,\partial_\theta,\partial_\varphi\}$. This decomposition is not unique, allowing for spatial rotations and Lorentz boosts. A possible choice is given by:
\begin{subequations}
\begin{align}
\hat{e}_{(\theta)}=A^\theta\partial_\theta,&\qquad \hat{e}_{(r)}=A^r\partial_r,\\
\hat{e}_{(\varphi)}=A^\varphi\partial_\varphi,&\qquad\hat{e}_{(t)}=\zeta\,\partial_t+\gamma\,\partial_\varphi,
\end{align}
\end{subequations}
where $\{\zeta,\gamma,A^r,A^\theta,A^\varphi\}$ are real coefficients. This particular choice is connected to a reference frame with zero axial angular momentum in relation to spatial infinity, and hence it is sometimes called the ZAMO reference frame, standing for \textit{zero angular momentum observers}\cite{Novikov}. An observer at rest in this frame moves with respect to the coordinate system, as a consequence of frame-dragging. The observer basis has a Minkowski normalization:
\begin{subequations}
\begin{align}
1=\hat{e}_{(\theta)}\cdot\,\hat{e}_{(\theta)},\quad&\qquad1=\hat{e}_{(\varphi)}\cdot\,\hat{e}_{(\varphi)},\\
1=\hat{e}_{(r)}\cdot\,\hat{e}_{(r)},\quad&\qquad {-1}=\hat{e}_{(t)}\cdot\,\hat{e}_{(t)}.
\end{align}
\end{subequations}
We also require that $0=\hat{e}_{(t)}\cdot\,\hat{e}_{(\varphi)}$. Using these conditions we obtain:
\begin{equation}
A^\theta=\frac{1}{\sqrt{g_{\theta\theta}}},\qquad A^r=\frac{1}{\sqrt{g_{rr}}},\qquad A^\varphi=\frac{1}{\sqrt{g_{\varphi\varphi}}},
\end{equation}
where the sign of the square roots was chosen positive so that at spatial infinity we have the standard orthogonal basis in spherical coordinates. Similarly we get \cite{Johannsen, Bardeen}:
\begin{equation}
\gamma=-\frac{g_{t\varphi}}{g_{\varphi\varphi}}\sqrt{\frac{g_{\varphi\varphi}}{g_{t\varphi}^2-g_{tt}g_{\varphi\varphi}}},\qquad\qquad\zeta=\sqrt{\frac{g_{\varphi\varphi}}{g_{t\varphi}^2-g_{tt}g_{\varphi\varphi}}}.
\label{zeta eq}
\end{equation}

A local measurement of a particle's property is performed by an observer at a given frame in the same position as the particle. Thus, the locally measured energy $p^{(t)}$ of a photon is given by the projection of its 4-momentum $p^\mu$ onto $\hat{e}_{(t)}$ \cite{Bardeen}:
\begin{equation}p^{(t)}=-(\hat{e}_{(t)}^\mu\, p_\mu)=-(\zeta p_{t}+\gamma p_\varphi).\end{equation}
The minus sign is a consequence of the time-like normalization $\hat{e}_{(t)}\cdot\,\hat{e}_{(t)}=-1$.
The quantities $E\equiv-p_t$ and $p_\varphi\equiv\Phi$ are conserved due to the associated Killing vectors, and they turn out to be respectively the photon's energy and angular momentum relative to a static observer at spatial infinity\footnote{This statement can be justified in the limit $r\to\infty$, which leads to $ p^{(t)}=E$ and to $p^{(\varphi)}r\sin\theta=\Phi$.} \cite{Bardeen} .\\
The locally measured linear momentum of the photon in all three spatial directions is obtained similarly, and so we obtain overall:
\begin{equation}\boxed{p^{(t)}=E\zeta-\gamma\Phi,}\qquad\qquad\boxed{p^{(\theta)}=\hat{e}_{(\theta)}^\mu \,p_\mu = \frac{1}{\sqrt{g_{\theta\theta}}}p_{\theta},}\label{locally measured theta momentum}
\end{equation}
\begin{equation}
\boxed{p^{(\varphi)}=\hat{e}_{(\varphi)}^\mu \,p_\mu = \frac{1}{\sqrt{g_{\varphi\varphi}}}\Phi,}\qquad\quad\boxed{p^{(r)}=\hat{e}_{(r)}^\mu \,p_\mu = \frac{1}{\sqrt{g_{rr}}}p_{r}.}\label{locally measured r momentum}
\end{equation}
A photon with zero angular momentum ($\Phi=0$) is observed in the ZAMO frame with no momentum component in the $\hat{e}_{(\varphi)}$ direction. This is due to the fact that an observer at rest at ZAMO also has zero angular momentum with respect to infinity, as was previously stated.

\subsection{Impact parameters $x,y$}

\begin{figure}[ht]
\begin{center}
\includegraphics[width=11cm]{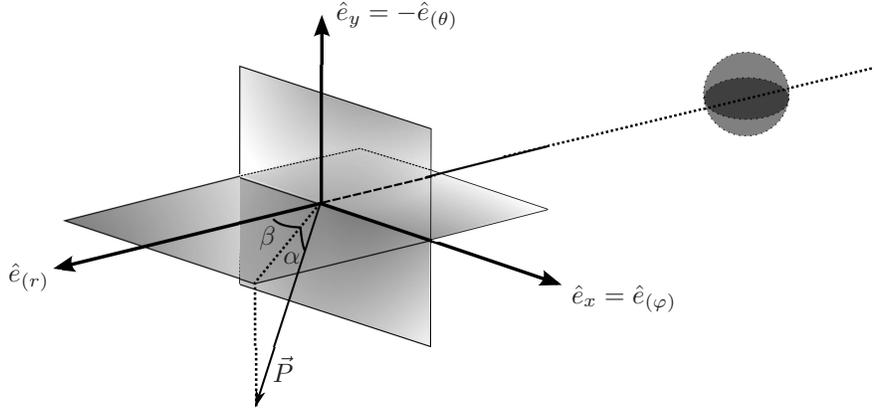}
\put(-206,147){$\hat{e}_{y}=-\hat{e}_{(\theta)}$}
\put(-117,40){$\hat{e}_{x}=\hat{e}_{(\varphi)}$}
\put(-330,48){$\hat{e}_{(r)}$}
\put(-226,55){$\alpha$}
\put(-235,62){$\beta$}
\put(-230,10){$\vec{P}$}
\end{center}
\vspace*{2pt}
\caption{\footnotesize Perspective drawing of the geometric projection of the photon's linear momentum $\vec{P}$ in the observer's frame $\{\hat{e}_{(r)},\hat{e}_{(\theta)},\hat{e}_{(\varphi)}\}$. The observation angles $\alpha,\beta$ were drawn as positive. The planes associated with the angles $\alpha$ and $\beta$ are perpendicular between themselves and the 3-vector $\vec{P}$ is in the same plane as $\alpha$. The vectors $\hat{e}_{x},\hat{e}_{(r)}$ and also coplanar with $\beta$. The BH is represented by the grey sphere in the image. 
}
\label{graph_angles_alfa_beta}
\end{figure}

\begin{figure}[ht]
\begin{center}
\includegraphics[width=10cm]{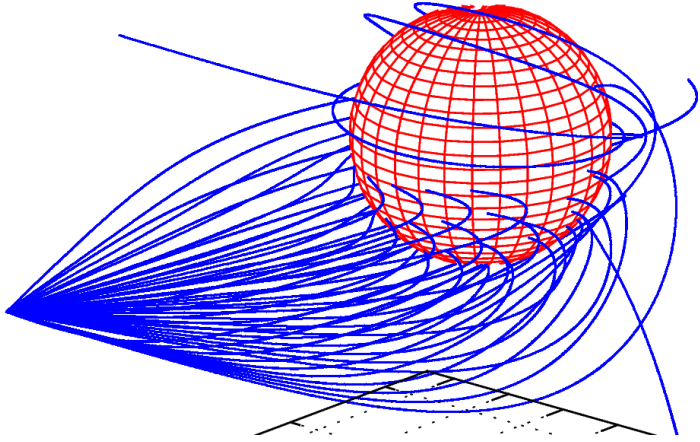}
\end{center}
\vspace*{2pt}
\caption{\footnotesize Graphical representation of the light rays (blue lines) and a Kerr BH (red sphere) with spin $a=0.8M$. The numerical data is displayed as if the Boyer-Lindquist coordinates were spherical. The rays branching point in the figure corresponds to the position of the observer, which is on the equatorial plane with $\tilde{r}=15M$. {The red sphere corresponds to the location of the event horizon.}
}
\label{graph_rays}
\end{figure}

Consider the projection of photons detected in an image plane, corresponding to the optical perspective of an observer. The latter could be taken as a camera. The Cartesian coordinates ($x,y$) assigned to each photon in this image plane are its impact parameters \cite{Johannsen} and they are proportional to the respective observation angles $(\alpha,\beta)$ (see Fig. \ref{graph_angles_alfa_beta}).\\
The solid angle that a given object occupies in the observer's sky {is a well defined concept and it depends strongly on the distance between the the object and the observer. However, this ``distance'' can be a very subtle concept in a curved space-time.} For instance, the proper distance to a Kerr BH's event horizon can diverge (in the extremal case) \cite{Bardeen}. For this reason the perimetral radius was used instead. Given a circumference at the equator ($\theta=\pi/2$) with constant radial coordinate $r$, its perimeter $\cal{P}$ is obtained by: 
\begin{equation}\mathcal{P}=\int^{2\pi}_0 \sqrt{g_{\varphi\varphi}}\,d\varphi=2\pi\sqrt{g_{\varphi\varphi}}.\end{equation}
Since $g_{\varphi\varphi}$ does not have a dependence on the coordinate $\varphi$ the integration was trivial. The perimetral or circumferential radius $\tilde{r}$ is then defined as:
\begin{equation}\tilde{r}\equiv\frac{\mathcal{P}}{2\pi}=\sqrt{g_{\varphi\varphi}}.\label{perimetral_radius}\end{equation}
This quantity is a possible choice to measure the distance to a BH. We expect the observation angles $\alpha,\beta$ to have a $1/\tilde{r}$ dependence as we approach spatial infinity. Hence, the impact parameters can be naturally defined as\cite{Johannsen,Bardeen}:
\begin{equation}x\equiv-\tilde{r}\beta\hspace{1cm}\textrm{and}\hspace{1cm}y\equiv\tilde{r}\alpha,\label{xy_parameters}\end{equation}
where the perimetral radius $\tilde{r}$ is computed at the position of the observer. The minus sign in the $x$ definition comes from the sign convention for $\beta$ (see Fig. \ref{graph_angles_alfa_beta}). Notice the observation angles $(\alpha,\beta)$ are both zero in the direction pointing to the center of the BH, in the observer's frame. Furthermore, the vectors $\hat{e}_{x}$ and $\hat{e}_{y}$ that span the image plane are defined by:
\begin{equation}\hat{e}_{x}=\hat{e}_{(\varphi)}\hspace{1cm}\textrm{and}\hspace{1cm}\hat{e}_{y}=-\hat{e}_{(\theta)}.\end{equation}
The 3-vector $\vec{P}$ in Fig. \ref{graph_angles_alfa_beta} is the photon's linear momentum with components $p^{(r)},p^{(\theta)}$ and $p^{(\varphi)}$ in the orthonormal basis $\{\hat{e}_{(r)},\hat{e}_{(\theta)},\hat{e}_{(\varphi)}\}$. We then have:
\begin{equation}|\vec{P}|^2=\left[p^{(r)}\right]^2+\left[p^{(\theta)}\right]^2+\left[p^{(\varphi)}\right]^2.\end{equation}
Moreover, attending to the geometry of the photon's detection (see Fig. \ref{graph_angles_alfa_beta}), we obtain:
\begin{subequations}
\begin{align}
&p^{(\varphi)}=|\vec{P}|\sin\beta\,\cos\alpha,\\
&p^{(\theta)}=|\vec{P}|\sin\alpha,\\
&p^{(r)}=|\vec{P}|\cos\beta\,\cos\alpha.
\end{align}
\label{p_decomposition}
\end{subequations}\\
Since the photon has {zero} mass $|\vec{P}|=p^{(t)}$.
The angular coordinates $(\alpha,\beta)$ (see Fig. \ref{graph_angles_alfa_beta}) of a point in the observer's local sky define the direction of the associated light ray and establishes its initial conditions. Combining the 4-momentum projections (\ref{locally measured theta momentum} - \ref{locally measured r momentum}) and the linear momentum $\vec{P}$ decomposition (\ref{p_decomposition}), we obtain:\\
\begin{subequations}
\label{initial_conditions}
\begin{align}
&p_\theta=|\vec{P}|\,\sqrt{g_{\theta\theta}}\sin\alpha,\qquad\qquad\,\,\,\Phi=|\vec{P}|\,\sqrt{g_{\varphi\varphi}}\sin\beta\,\cos\alpha,\\
&p_r=|\vec{P}|\,\sqrt{g_{rr}}\cos\beta\,\cos\alpha,\qquad E=|\vec{P}|\,\left(\frac{1+\gamma\sqrt{g_{\varphi\varphi}}\sin\beta\,\cos\alpha}{\zeta}\right).
\end{align}
\end{subequations}\\
Curiously the value of $|\vec{P}|$ is redundant for the geodesic trajectory since its variation leads to a simple rescaling of the affine parameter. In fact, $|\vec{P}|$ only establishes the photon's frequency and does not influence the trajectory itself. For this reason this value can be set to unity for simplicity.\\
In practice the observer's local sky is divided into small solid angles, each corresponding to a pixel in the final image (see Fig. \ref{graph_rays}). For each pixel the initial conditions are determined from its coordinates $(\alpha,\beta)$. The geodesic equation is then integrated numerically backward, which amounts to a negative variation of the affine parameter.
If the photon reaches either a light source or {(asymptotically) approaches a} BH then the integration stops, there is optical information assigned to the respective pixel and the process is repeated.\\

\section{Kerr Space-time}

In Boyer-Lindquist coordinates $\{t,r,\theta,\varphi\}$ the Kerr metric is given by \cite{Novikov}:
\[ds^2=-\left(1-\frac{2Mr}{\rho^2}\right)dt^2 -\left(\frac{4Mar\sin^2\theta}{\rho^2}\right)dtd\varphi +\left(\frac{\rho^2}{\Delta}\right)dr^2 + \rho^2d\theta^2 +\]\begin{equation}+\sin^2\theta\left(r^2+a^2+{\frac{2Mra^2\sin^2\theta}{\rho^2}}\right)d\varphi^2,\label{Kerr metric}\end{equation}
where $\rho^2=r^2+a^2\cos^2\theta$ and $\Delta=r^2-2Mr+a^2$. This metric depends only on two parameters: the ADM mass $M$ of the BH and the rotation parameter $a\in[-M,M]$, the latter proportional to the ADM angular momentum\footnote{More precisely, one cannot define ADM angular momentum in the same way as ADM mass \cite{Gourgoulhon}, and the quantity we are dubbing as ADM angular momentum is the Komar angular momentum at spatial infinity.}.\\
Using the Hamilton-Jacobi formalism \cite{Carter} it is possible to write the geodesic equations in Kerr space-time as four first-order differential equations:
\begin{subequations}
\begin{align}
\rho^2\dot{r}&=\pm\sqrt{\mathcal{R}}, \qquad\mathrm{with}\qquad \mathcal{R}\equiv H^2-\Delta[Q+(aE-\Phi)^2+\mu^2r^2],\\
\rho^2\dot{\theta}&=\pm\sqrt{\Theta}, \qquad\textrm{with}\qquad\Theta=Q-\cos^2\theta\left(a^2(\mu^2-E^2)+\frac{\Phi^2}{\sin^2\theta}\right).\\
\rho^2\dot{t}&=\frac{E}{\Delta}[(r^2+a^2)^2-a^2\Delta\sin^2\theta]-\frac{2Mar}{\Delta}\Phi.\\
\rho^2\dot{\varphi}&=\frac{2MaEr}{\Delta}+\Phi\frac{(\Delta-a^2\sin^2\theta)}{\Delta\sin^2\theta},
\end{align}
\label{Kerr_geodesic}
\end{subequations}
where $H\equiv E(r^2+a^2)-a\Phi$. This simplification is possible due to the existence of four constants of motion $\{E,\Phi,Q,\mu\}$, where the rest mass $\mu=0$ for the photon and $Q$ is the Carter constant. The latter is a consequence of a hidden symmetry of Kerr.\\
In Kerr space-time there are unstable photon orbits with constant radial coordinate, dubbed ``spherical photon orbits''. The value of the reduced constants $\lambda\equiv \Phi/E$ and $\eta\equiv Q/E^2$ are fixed for a given spherical orbit at a radial coordinate $RM\equiv r$:
\begin{equation}\lambda=-\frac{R^3-3R^2+a^2R+a^2}{a(R-1)},\qquad\eta=-\frac{R^3(R^3-6R^2+9R-4a^2)}{a^2(R-1)^2}.\end{equation}
where $R\in[r_1,r_2]$. These are defined as the roots\cite{Chandra,Bardeen} of $\eta$:

\begin{subequations}
\begin{align}
&r_1=2\left\{1+\cos\left(\frac{2}{3}\arccos\left[-\frac{|a|}{M}\right]\right)\right\},\\
&r_2=2\left\{1+\cos\left(\frac{2}{3}\arccos\left[\frac{|a|}{M}\right]\right)\right\}.
\end{align}
\end{subequations}
 Spherical orbits are the closest photons {coming from infinity} can get to the BH and still be able to escape. The shadow edge is thus created by photons that almost follow spherical orbits and, because these are unstable, they are on the verge of either being captured or escaping.

\subsection{Analytical form of the Kerr shadow }

\begin{figure}[ht]
\begin{center}
\includegraphics[width=10cm]{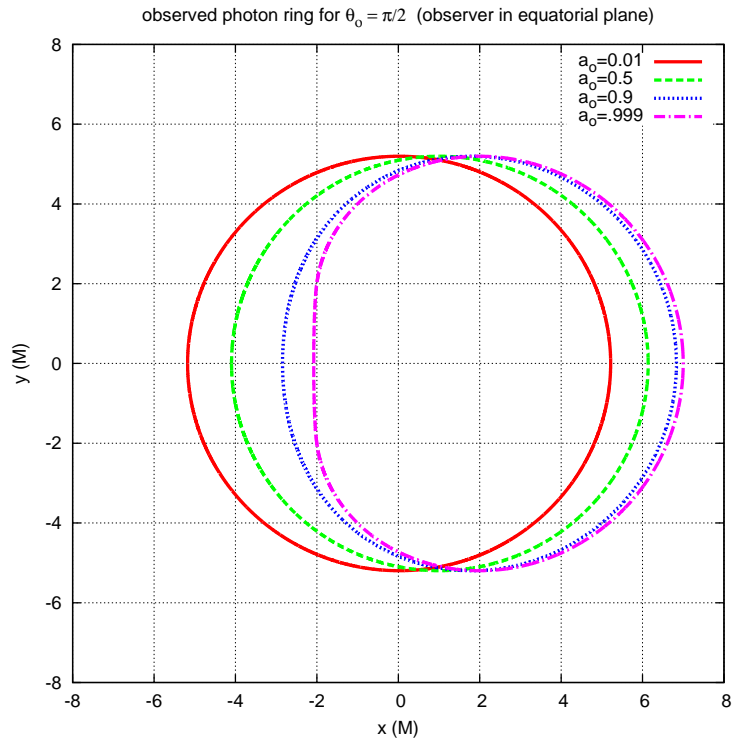}
\end{center}
\vspace*{2pt}
\caption{\footnotesize Representation of the Kerr shadow rim from the analytical solution, as observed from very large distances ($r_o\gg M$) in the equatorial plane of the BH $(\theta_o=\pi/2)$. Different values of the dimensionless rotation parameter $a_o=a/M$ are displayed. Notice that for $a_o\simeq0$ we have almost a circle due to the symmetry of the Schwarzschild solution and for $a_o\simeq 1$ we have a highly asymmetric geometric shape due to the symmetry violation in the latitude coordinate $\theta$ for the Kerr metric.
}
\label{graph_Kerr_analy}
\end{figure}

The shadow's edge of a Kerr BH can be calculated in an analytical closed form. In the following calculation the observer is at a ZAMO frame at radial coordinate $r_o$ and latitude coordinate $\theta_o$. Starting from (\ref{p_decomposition}) and solving for the observation angles $\alpha,\beta$ we obtain:
\begin{equation}\tan\beta=\frac{p^{(\varphi)}}{p^{(r)}},\qquad \sin\alpha=\frac{p^{(\theta)}}{p^{(t)}}.\end{equation}
For an observer facing the BH, photons coming from the shadow edge have $p^{(r)}\geq 0$, and so we have $p^{(r)}\geq 0\implies \cos\beta\geq 0$. Since the domain of $\alpha$ is $[-\pi/2:\pi/2]$ we obtain:
\begin{equation}\beta=\arctan\left[\frac{p^{(\varphi)}}{p^{(r)}}\right],\qquad\alpha=\arcsin\left[\frac{p^{(\theta)}}{p^{(t)}}\right].\end{equation}
Combination of (\ref{locally measured theta momentum}), (\ref{locally measured r momentum}) and (\ref{Kerr_geodesic})  yields:
\begin{subequations}
\begin{align}
p^{(\theta)}=\pm\frac{\sqrt{\Theta}}{\sqrt{g_{\theta\theta}}},&\qquad p^{(\varphi)}=\frac{\Phi}{\sqrt{g_{\varphi\varphi}}},\\
p^{(r)}=\frac{\sqrt{R}}{\Delta\sqrt{g_{rr}}},&\qquad p^{(t)}=E\zeta-\gamma\Phi.
\end{align}
\end{subequations}
Using the definition of the reduced photon constants $\lambda\equiv\Phi/E$ and $\eta\equiv Q/E^2$ and of the impact parameters $x,y$:
\begin{equation}y=\tilde{r}\arcsin\left[\frac{\pm1}{(\zeta-\lambda\gamma)}\frac{\sqrt{\eta+a^2\cos^2\theta_o-\lambda^2/\tan^2\theta_o}}{\sqrt{r_o^2+a^2\cos^2\theta_o}}\right],\end{equation}
\begin{equation}
x=-\tilde{r}\arctan\left[\frac{\lambda\sqrt{\rho^2\Delta}}{\sqrt{g_{\varphi\varphi}}\sqrt{r_o^4+(a^2-\eta-\lambda^2)r_o^2+2mr_o[\eta+(a-\lambda)^2]-\eta a^2}}\right].
\end{equation}
In the previous expressions the impact parameters $\lambda,\eta$ are functions of the spherical photon orbit coordinate radius. The rim of the shadow's edge in the $x,y$ image plane is then defined parametrically, as $R$ changes in the interval $R\in[r_1,r_2]$. An example is displayed in Fig. \ref{graph_Kerr_analy}. 
\subsection{Numerical Kerr shadow}
\begin{figure}[ht]
\begin{center}
\begin{tabular}{c}
\includegraphics[width=7cm]{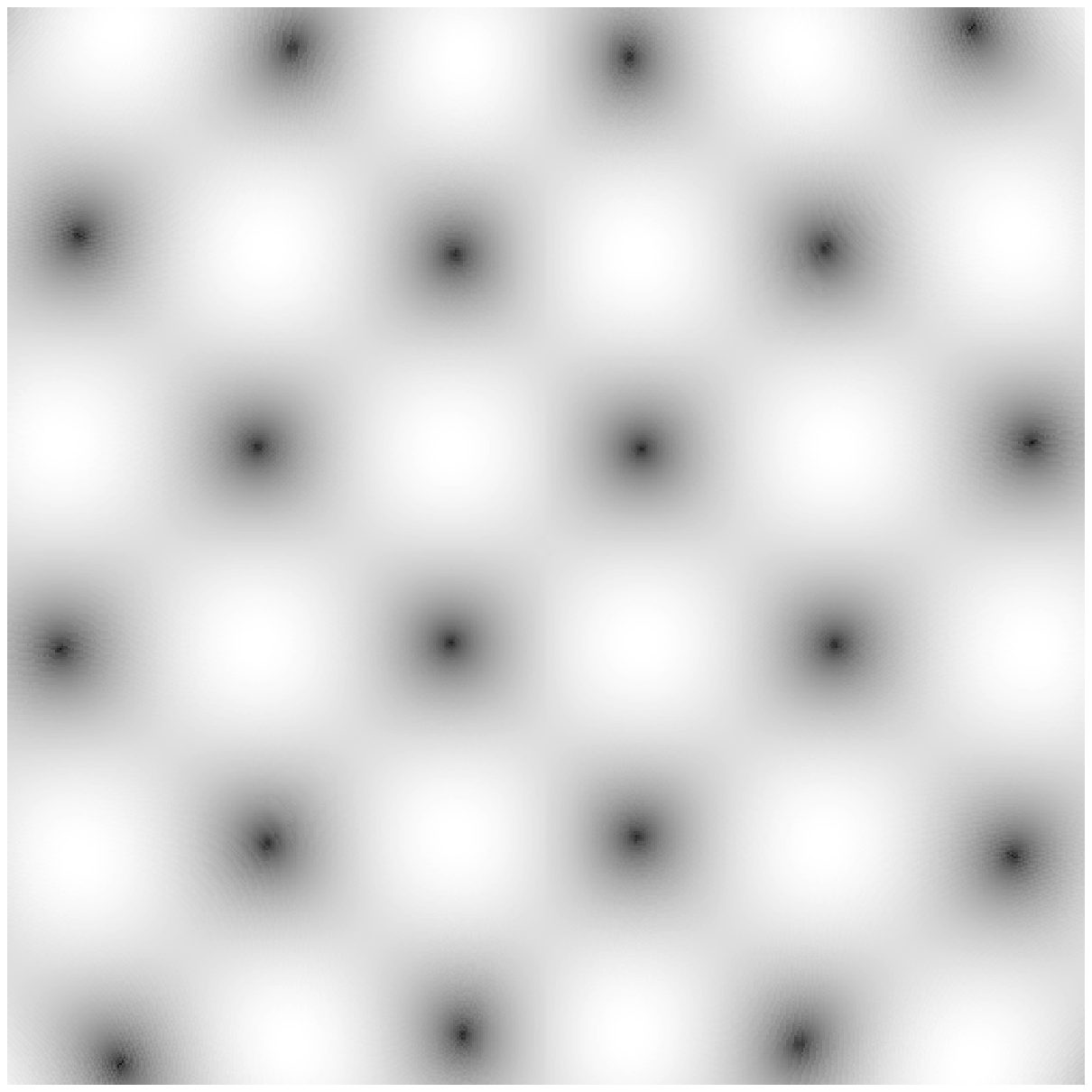}\\
\includegraphics[width=7cm]{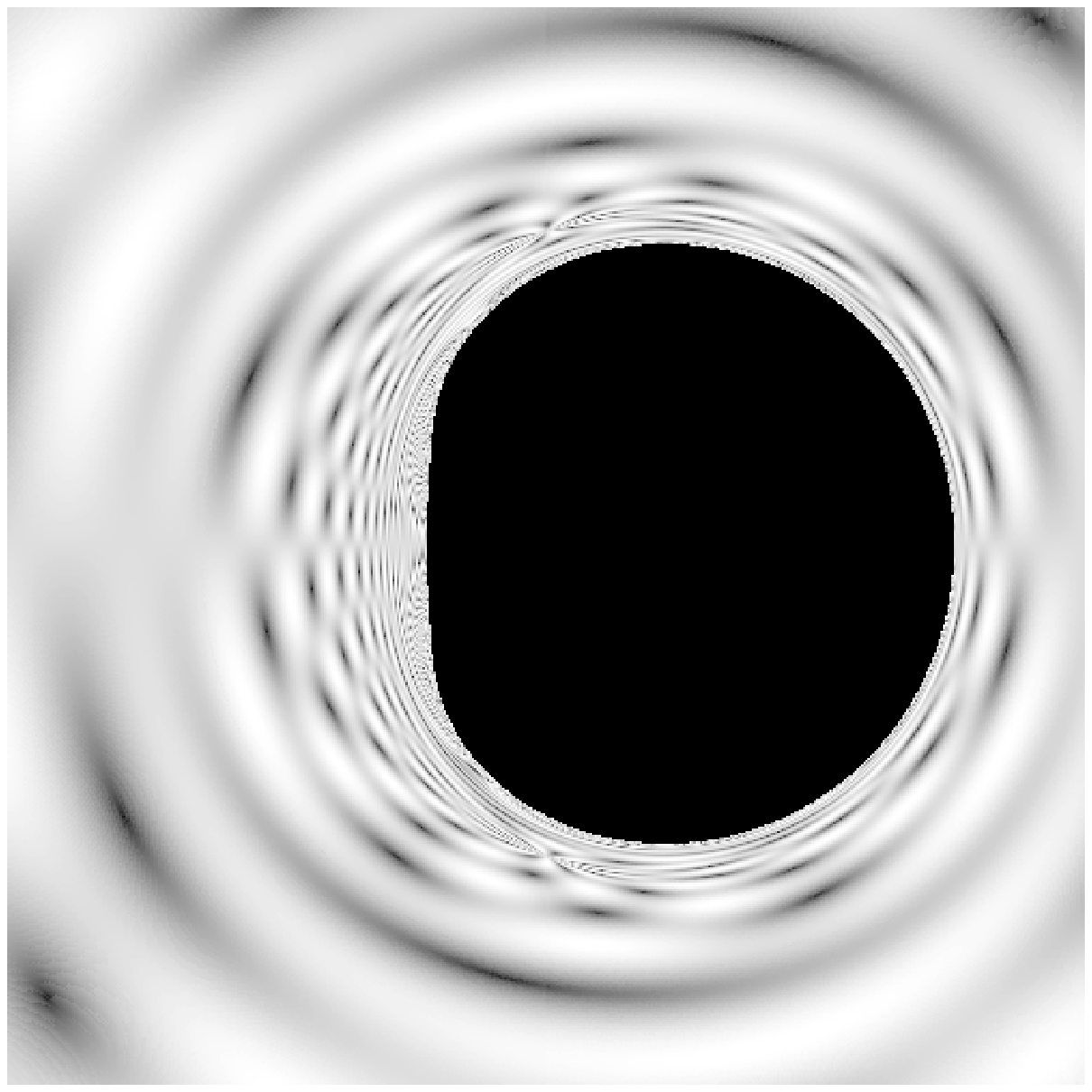}
\end{tabular}
\end{center}
\vspace*{2pt}
\caption{\footnotesize (\textit{Top}:) Pattern imprinted on the celestial sphere, in flat space. (\textit{Bottom}:) Shadow of an extremal Kerr BH ($a=M$) obtained numerically by backward ray-tracing. In both images the observer is in the equatorial plane with $\tilde{r}=15M$ and has the same field of view. The celestial sphere is at $\tilde{r}=30M$ in both cases.
}
\label{graph_Kerr}
\end{figure}

Despite the existence of an analytical solution, the Kerr shadow can still be obtained numerically as a cross-check. To illustrate the shadow and the gravitational lensing, a large celestial sphere acting as a light source was placed at $\tilde{r}=30M$. The latter englobes and encloses both the BH and the observer, but it is concentric with the first. The gravitational lensing is thus a mapping between the angles $\alpha,\beta$ and a point in the celestial sphere. If a pattern is ``imprinted" on the celestial sphere (see Fig. \ref{graph_Kerr} for instance), then a ray that reaches a given point on that surface retrieves the pattern information at that point back to the initial pixel. On the other hand, a ray that {originates (asymptotically)} from a point on the event horizon of the BH corresponds naturally to a black pixel (no light came from that direction).\\
Although two BHs with different values of $a$ are each still a member of the Kerr family, they have distinct space-time geometries; the coordinate $r$ is not even directly comparable between such solutions. At this point it is important to establish a criterion for similar observation conditions in different space-times: two observers are in similar observation conditions if the perimetral radius $\tilde{r}$ is the same for both observers. This implies 
\begin{equation}\sqrt{{g_{\varphi\varphi}}^{(1)}}=\sqrt{{g_{\varphi\varphi}}^{(2)}},\end{equation}
where each superscript (1) and (2) labels the respective space-time. Still, some reference distance must be provided to make the final link. The scale defined by the ADM mass $M$ will be used for such a purpose. An interesting point is that for flat space in spherical coordinates (which is a limiting case of the Kerr family) we have: 
\begin{equation}\tilde{r}=\sqrt{{g_{\varphi\varphi}}^{\textrm{(flat)}}}=r^{\textrm{(flat)}}.\end{equation} 
Comparing observations in flat space with observations for a Kerr BH at a radial coordinate $r$ yields then:
\begin{equation}\tilde{r}=\sqrt{r^2+a^2+\frac{2Ma^2}{r}}.\end{equation}
 The inversion of this equation leads to:
\begin{equation}r=2\sqrt{\frac{\tilde{r}^2-a^2}{3}}\cos\left(\frac{1}{3}\arccos\left[\frac{3a^2M}{a^2-\tilde{r}^2}\sqrt{\frac{3}{\tilde{r}^2-a^2}}\right]\right).\label{conversion_r_perimetral}\end{equation}
So, given a radius $\tilde{r}$ in flat space, we can compute the equivalent radial coordinate $r$ in Kerr space-time that leads to similar observation conditions. In practice however, the difference $(\tilde{r}-r)$ is  quite small compared with $\tilde{r}$, unless the Kerr observer is very close to the BH. For example, for $a=0.999M$ we have $\tilde{r}=15M\implies r\simeq 14.96M$.\\

\section{Kerr BHs with Scalar Hair}

Consider a complex massive scalar field $\phi$ minimally coupled to Einstein's gravity. The action $\mathcal{S}[g_{\mu\nu},\phi]$ is given by:
\begin{equation}\mathcal{S}[g_{\mu\nu},\phi]=\int d^4x\sqrt{-g}\left[\frac{R}{16\pi}-\nabla_\mu\phi\nabla^\mu\phi^*-\eta^2\phi^*\phi\right],\end{equation}
where $g$ is the determinant of the metric, $R$ is the Ricci scalar and $\eta$ is the mass of the scalar particle. Besides the Einstein field equations, the variational principle yields the massive Klein-Gordon equation for the field: $\nabla_\mu\nabla^\mu\phi=\eta^2\phi$. It is possible to find a family of BH solutions in equilibrium with the scalar field, dubbed in literature as \textit{Kerr BHs with scalar hair} \cite{Herdeiro_PRL,Herdeiro_review}. This family has the Kerr space-time and rotating Boson Stars as limiting cases.\\
The ansatz for the space-time of these hairy BHs (HBHs) is assumed to be stationary and axially symmetric:
\begin{equation}ds^2=e^{2F_1}\left(\frac{dr^2}{N}+r^2d\theta^2\right)+e^{2F_2}r^2\sin^2\theta(d\varphi-Wdt)^2-e^{2F_0}Ndt^2,\end{equation}
where $N=1-r_H/r$ and $r_H$ is the radial coordinate of the BH event horizon. The ansatz for the scalar field is given by:
\begin{equation}\phi=\widetilde{\phi}(r,\theta)e^{i(m\varphi-wt)},\end{equation}
where $w$ is the field frequency and $m$ is an integer named azimuthal harmonic index. The functions $\widetilde{\phi},F_0,F_1,F_2,W$ are only known numerically \cite{Herdeiro_review}. \\
Both Kerr and HBHs are stationary and axially symmetric. However, the existence of a Carter constant $Q$ is specific for Kerr, consequence of an hidden symmetry (namely a Killing tensor). Since such a symmetry does not exist for HBHs, is not possible to reduce all the four geodesic equations to first order.

\begin{figure}[ht]
\begin{center}
\includegraphics[width=12cm]{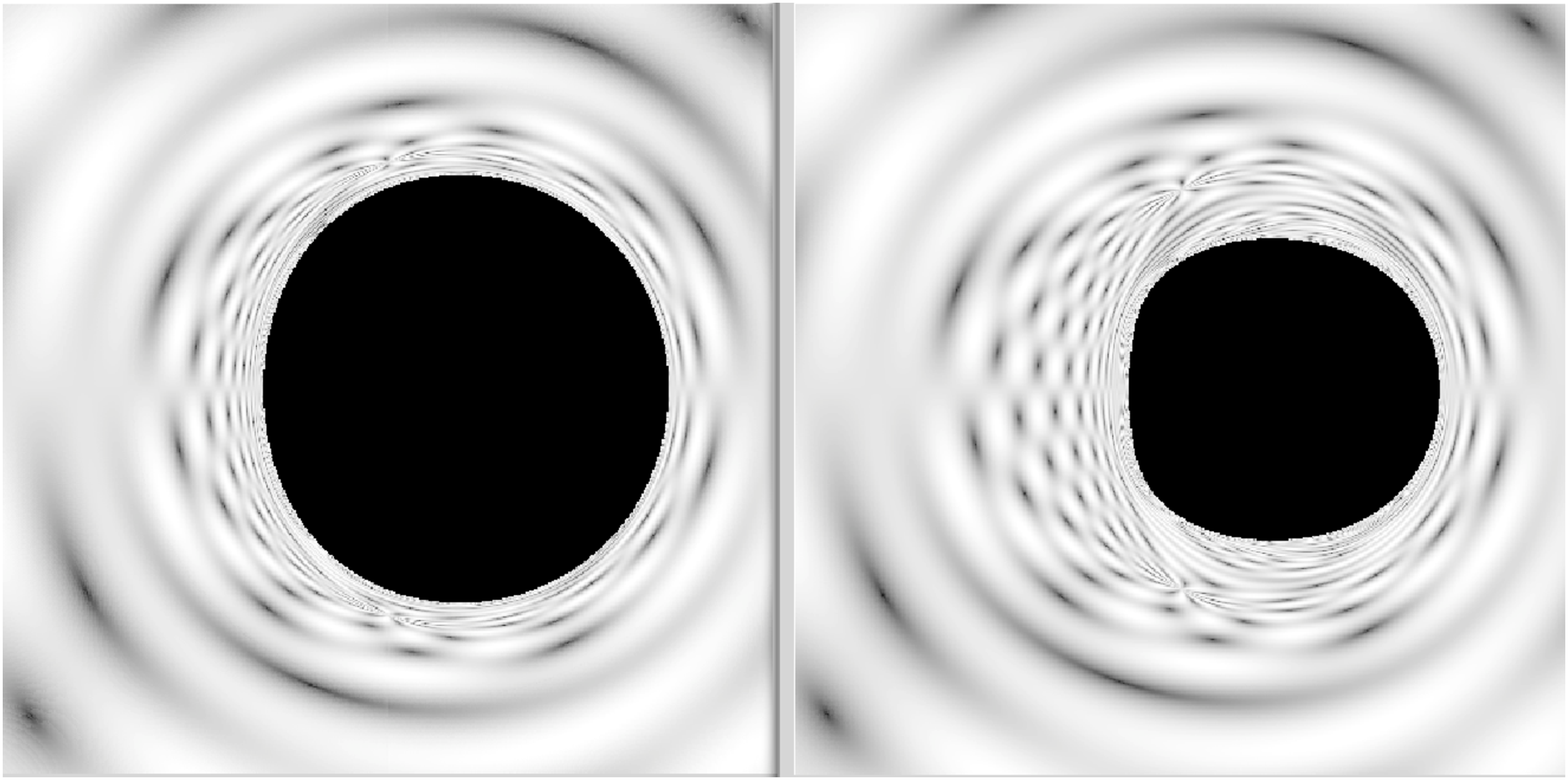}
\end{center}
\vspace*{2pt}
\caption{\footnotesize (\textit{Left}): Shadow of a Kerr BH with $a=0.849M$. (\textit{Right}): Shadow of a Kerr BH with scalar hair with the same ADM mass and angular momentum. The setup of both images is the same as in Fig.\ref{graph_Kerr}.
}
\label{graph_IV}
\end{figure}

\begin{figure}[ht]
\begin{center}
\begin{tabular}{c}
\includegraphics[width=7cm]{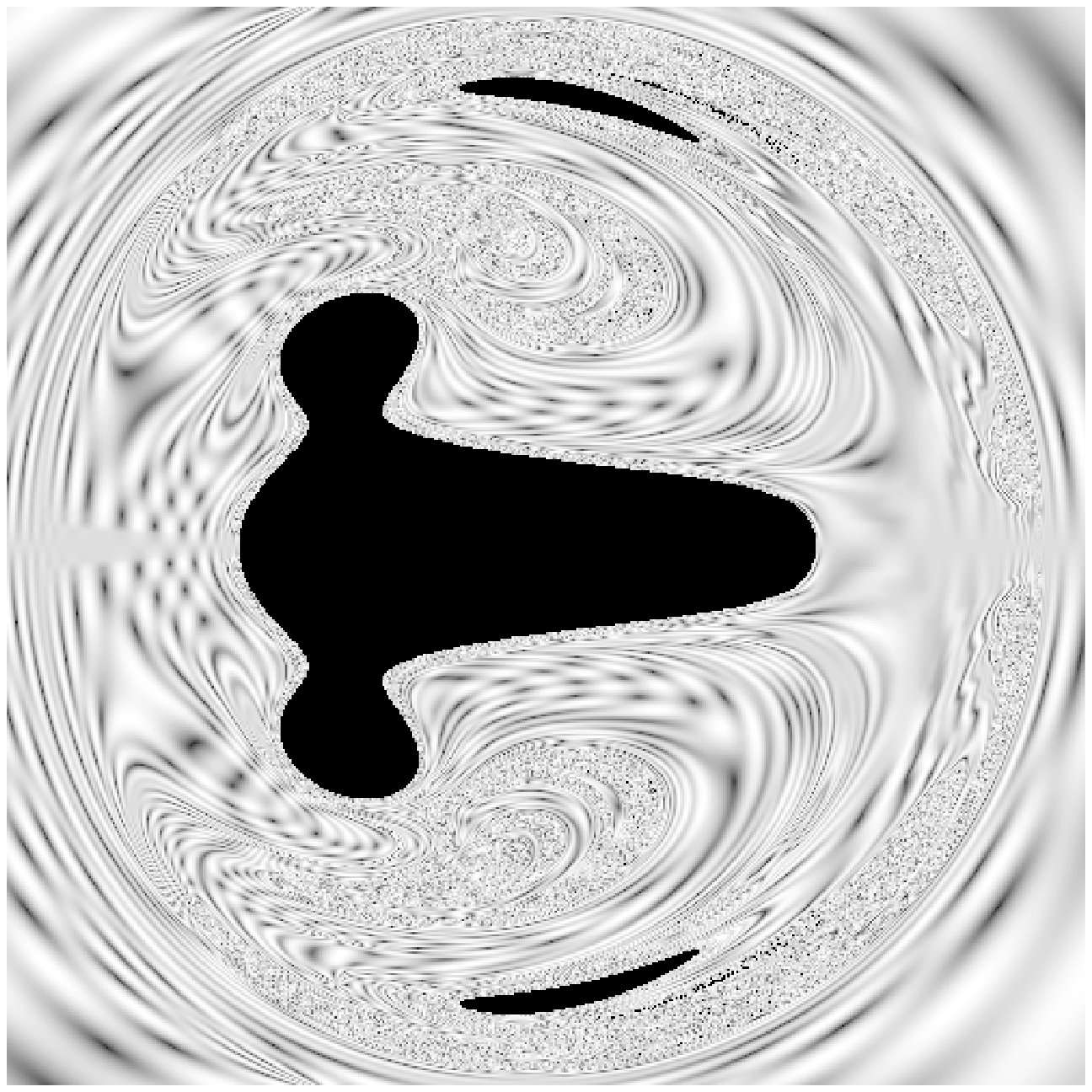}\\
\includegraphics[width=8.5cm]{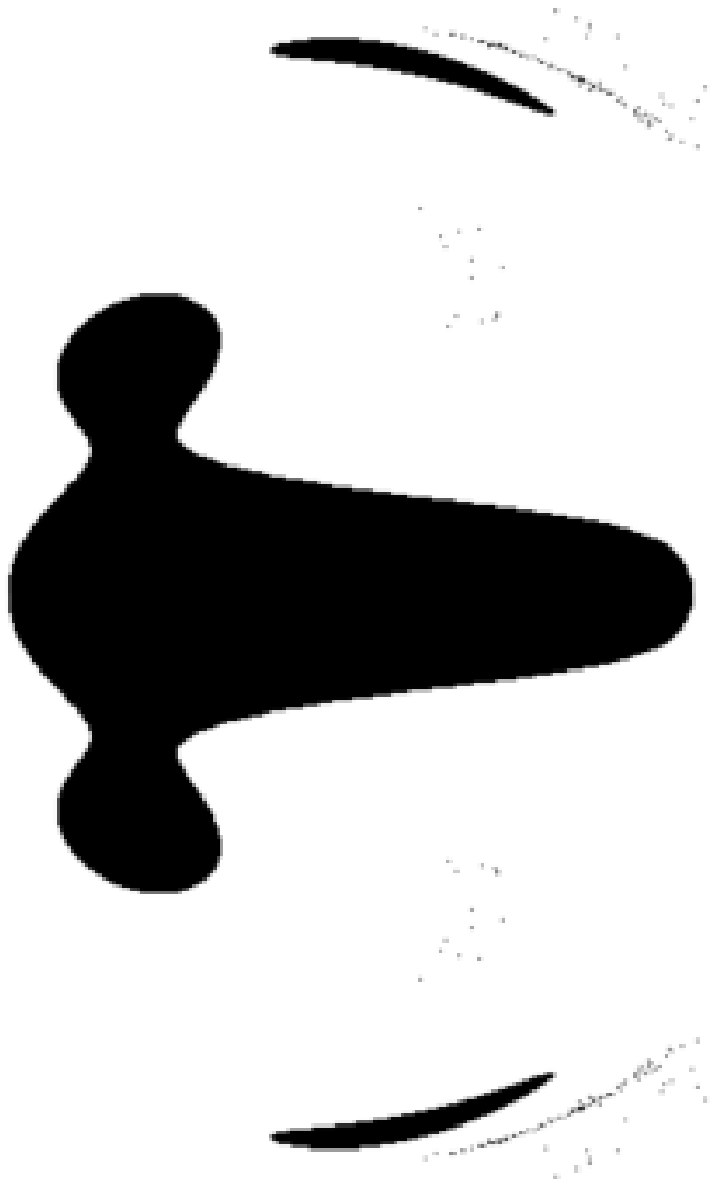}
\end{tabular}
\end{center}
\caption{\footnotesize (\textit{Top}:) Shadow example of a Kerr BH with scalar hair which violates the Kerr bound ($a^2>M^2$) in terms of horizon quantities. (\textit{Bottom}:) Representation of the shadow only, highlighting the fact that it is disconnected and that it has numerous smaller \textit{eyebrows}. The setup is essentially the same as in Fig.\ref{graph_Kerr}, except for a zoomed in field of view. 
}
\label{graph_V}
\end{figure}

In order to compare observations in Minkowski, Kerr, and HBH space-times a similar observation criterion is necessary. Different radial coordinates are equivalent if the perimetral radius $\tilde{r}$ is the same in the equatorial plane ($\theta=\pi/2$):
\begin{equation}\tilde{r}=\sqrt{r_k^2+a^2+\frac{2Ma^2}{r_k}}=r_q\,e^{F_2(r_q,\,\pi/2)},\end{equation}
where $r_k$ and $r_q$ are respectively the radial coordinates for Kerr and hairy BHs. Hence, given a value of $\tilde{r}$, one can obtain the equivalent radial coordinates between a Kerr BH and a HBH. As in Kerr, the HBH length scale can be provided by the ADM mass, although the usage of other quantities is also possible.\\
As an illustrative example, the comparison between a Kerr BH and a HBH with the same ADM mass and angular momentum is displayed in Fig. \ref{graph_IV}. The most prominent feature of this HBH shadow is that it is smaller and more squared-like with respect to the Kerr analogue. A more extreme case is given in Fig. \ref{graph_V}, for which the shadow becomes disconnected and exhibits a fairly exotic {profile}.\\

\section{Conclusions}

In order to obtain shadows of BHs numerically, the backward ray-tracing algorithm was described, including the construction of an observer frame. Additionally, an analytical solution for the Kerr shadow was also derived in closed form for an observer at an arbitrary position. Subsequently, numerical examples of shadows of Kerr BHs with scalar hair were displayed, contrasting with the Kerr analogue. Hence these HBHs could provide experimental templates for the upcoming observations with the Event Horizon Telescope. The latter aims to probe the supermassive BH candidate Sgr A* in the center of our galaxy.\\

\section*{Acknowledgments}

C.H. and E.R. acknowledge funding from the FCT-IF programme. P.C. is supported by Grant No. PD/BD/114071/2015 under the FCT-IDPASC Portugal Ph.D. program and by the Calouste Gulbenkian foundation. H.R. is supported by Grant No. PD/BD/109532/2015 under the MAP-Fis Ph.D. program. This work was partially supported by  the  H2020-MSCA-RISE-2015 Grant No.  StronGrHEP-690904, and by the CIDMA Project No. UID/MAT/04106/2013. Computations were performed at the Blafis cluster, in Aveiro University and at the Laboratory for Advanced Computing, University of Coimbra.\\


\end{document}